# The Effects of Chemical Doping and Hydrostatic Pressure on $T_c$ of $Y_{1-y}Ca_yBa_2Cu_3O_x$ and $NdBa_2Cu_3O_x$ Single Crystals

S.I. Schlachter[*], K.-P. Weiss[*+], W.H. Fietz[*], K. Grube[*],
H. Leibrock[*], Th. Wolf[*], B. Obst[*], P. Schweiss[‡], M. Kläser[+],
and H. Wühl[*+]

Forschungszentrum Karlsruhe, [*]ITP and [‡]INFP, 76021 Karlsruhe, Germany
[+]Universität Karlsruhe, IEKP, 76128 Karlsruhe, Germany

We studied $T_c$ of $YBa_2Cu_3O_x$ (Y123), $Y_{0.89}Ca_{0.11}Ba_2Cu_3O_x$ (YCa123) and $NdBa_2Cu_3O_x$ (Nd123) single crystals with various oxygen contents $x$. Compared to $T_c(x)$ of Y123 the $T_c(x)$ curves of YCa123 are shifted to lower oxygen contents and the maximum transition temperature $T_{c,max}$ decreases with increasing Ca content whereas in Nd123 $T_c(x)$ is shifted to higher oxygen contents and $T_{c,max}$ is increased. According to the universal parabolic $T_c(n_h)$ behavior the differences in $T_c(x)$ of Y123, YCa123 and Nd123 can be ascribed to different hole concentrations $n_h$ in the $CuO_2$ planes caused by doping via changes in chemistry or structure. In order to study the influence of structural changes on $T_c$ we examined the hydrostatic pressure effect $dT_c/dp$ ($p \leq 0.6 GPa$). In the underdoped region, at $n_h \approx 0.11$, the examined compounds show a peak in $dT_c/dp$ which is very pronounced for systems with well ordered CuO chains. As this peak occurs at the same $n_h$ in all investigated systems it is not caused by oxygen ordering, but its origins might be found in a strong influence of lattice deformations on the electronic structure.
PACS numbers: 74.62.-c, 74.62.Fj, 74.62.Dh, 74.72.-h, 74.72.BK

Presland *et al.*[1] and Tallon *et al.*[2] showed that many high temperature superconductors exhibit a universal parabolic $T_c(n_h)$ relationship:

$$\frac{T_c}{T_{c,max}} = 1 - \left(\frac{n_h - n_{h,opt}}{0.11}\right)^2 \qquad (1)$$

where $n_h$ is the hole concentration per $CuO_2$ plane and $T_{c,max}$ is the maximum transition temperature at optimum doping $n_{h,opt} \approx 0.16$. In the $RBa_2Cu_3O_x$ system (R123, R = Y, Ca or rare earth elements) $n_h$ can be changed in different ways, for example by variation of oxygen content, substitution of $Ca^{2+}$ for the $R^{3+}$ ion or by application of pressure. The first and second method introduce additional holes to the unit cell whereas application of pressure changes $n_h$ only by charge transfer between the CuO chains and the $CuO_2$ planes[3,4], leaving the chemical composition of a sample unchanged. Besides pressure-induced charge transfer there are other effects which change $T_c$ remarkably under pressure. According to Eq. (1) also a change of the maximum transition temperature and optimum hole concentration under pressure, $dT_{c,max}/dp$ and $dn_{h,opt}/dp$, respectively, should have a strong influence on $T_c$. In this work we studied $T_c$ and $dT_c/dp$ of Y123, YCa123 (11% Ca) and Nd123 single crystals with various oxygen contents $x$ and compared the dependences of $T_c$ and $dT_c/dp$ on the oxygen content $x$ to the dependence on $n_h$.

The preparation and characterization of the Y123 and YCa123 single crystals are described in Ref. 5. The Nd123 single crystals were synthesized in Y-stabilized $ZrO_2$ crucibles. Due to corrosion of the crucibles a small amount of Y is incorporated on Nd sites. Quantitative EDX analysis proofed the Y content to be approx. 8%. Neutron-diffraction studies did not show any Nd on Ba sites. The oxygen content of the Y123/YCa123 and Nd123 samples was estimated from isotherm curves $\log p_{O_2}$ vs. oxygen content from Ref. 6 and Ref. 7, respectively. In addition we checked the oxygen content of selected samples by neutron-diffraction studies and found the results to be in accordance with our estimations. AC-susceptibility measurements under hydrostatic He-gas pressure up to 0.6 GPa were performed in a CuBe pressure cell. In order to avoid pressure induced oxygen ordering effects[8-10] pressure was applied at low temperature ($T \leq 110$ K)[11] and the sample kept below 110 K during the whole experiment.

In Fig. 1a $T_c$ values of Y123, YCa123 and Nd123 are plotted versus oxygen content $x$. It can clearly be seen that $T_{c,max}$ of Y123 and YCa123 - 91.5 K and 86.3 K, respectively - are reached at different oxygen contents $x$. This is to be expected since in YCa123 additional charge carriers due to Ca doping enhance $n_h$ in the $CuO_2$ planes. Therefore, with increasing Ca content optimum doping $n_{h,opt}$ can already be achieved at lower oxygen contents. The optimum doping of Nd123 having the highest $T_{c,max} = 95.5$K of the examined systems seems to be reached at $x \approx 7.0$ and the onset of superconductivity is moved to higher oxygen contents than in Y123 (see also Ref. 12). According to Eq. (1) in each system the onset of superconductivity and the maximum transition temperature occur at $n_h = 0.05$ and $n_h = 0.16$, respectively. Therefore, for a distinct oxygen content the number of holes in the $CuO_2$ planes is not the same in Y123 and Nd123. The enlarged lattice parameters of Nd123 change the oxygen order in the CuO-chain subsystem which strongly influences the creation of holes in the unit cell. At $x = 6.5$ the oxygen order shows ortho-II[13] and herringbone[14] structure in Y123 and Nd123, respectively. With an ideal ortho-II structure 0.5 holes are created per formula unit[15] and distributed among CuO chains and $CuO_2$ planes. An ideal herringbone structure does not create any holes at all at $x = 6.5$. The different oxygen ordering processes are caused by attractive[16] and repulsive[17] interactions between chain-oxygen atoms in $b$-axis direction in Y123 and Nd123, respectively. In addition, the transfer of holes to the $CuO_2$ planes via the Ba and the apical-oxygen atom is influenced by the size of the $R^{3+}$ ion.

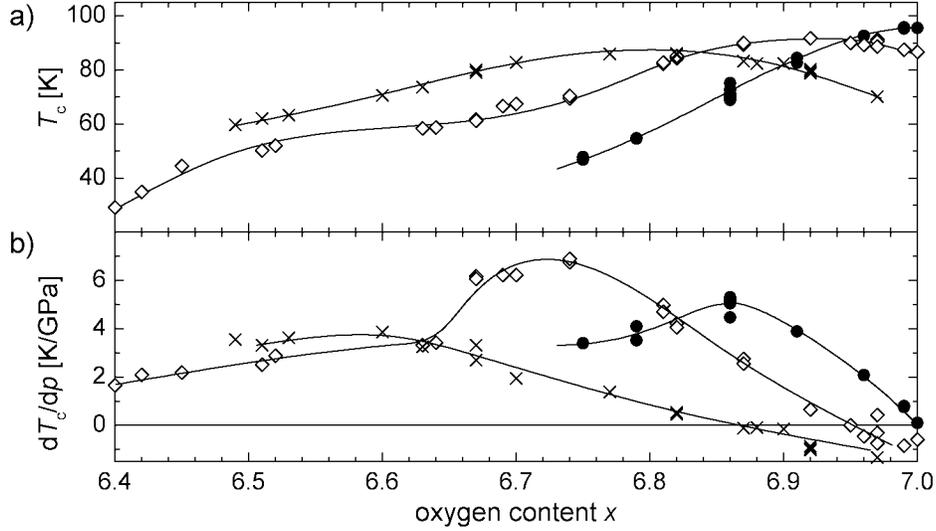

Fig. 1. $T_c$ and $dT_c/dp$ of $Y_{0.89}Ca_{0.11}Ba_2Cu_3O_x$ (×), $YBa_2Cu_3O_x$ (◇) and $NdBa_2Cu_3O_x$ (●) versus oxygen content $x$.

In Fig. 1b $dT_c/dp$ is plotted versus $x$. Underdoped samples of the Y123 and Nd123 systems both show a maximum in $dT_c/dp$ whereas in YCa123 the peak is extremely broadend and shows a nearly plateau-like behaviour below $x = 6.6$. With increasing oxygen content a zero crossing from positive to negative values slightly beyond $T_{c,max}$ can be seen in $dT_c/dp(x)$ of Y123 and YCa123. This means that the pressure effect at $T_{c,max}$ is small and positive ($dT_{c,max}/dp < 1K/GPa$). For Nd123 $dT_{c,max}/dp = 0.1K/GPa$, however, as the oxygen content can not be increased beyond $x = 7.0$ we were not able to detect a zero crossing. While for the three different systems the maxima as well as the zero crossings in $dT_c/dp$ occur at clearly different oxygen contents, the hole concentration for maximum pressure effect $n_{h,m}$ is the same in all three systems as well as the hole concentration for the zero crossing $n_{h,zc}$, as can be seen in Fig. 2, where $n_h$ was calculated from Eq. (1) using $T_c$ and $T_{c,max}$ values from Fig. 1a.

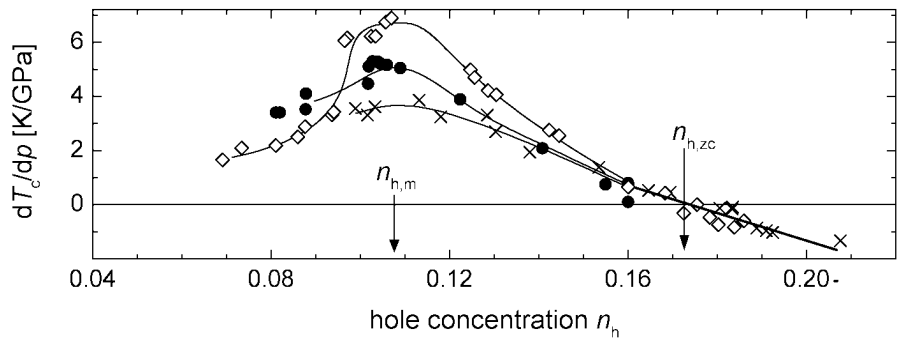

Fig. 2. $dT_c/dp$ of $Y_{0.89}Ca_{0.11}Ba_2Cu_3O_x$ ($\times$), $YBa_2Cu_3O_x$ ($\diamond$) and $NdBa_2Cu_3O_x$ ($\bullet$) versus hole concentration $n_h$.

In order to find the origins of $T_c$ variations under hydrostatic pressure it is useful to separate the effects along the three crystallographic axes: From thermal expansion[18,19] and uniaxial pressure[20] experiments together with theoretical calculations[21] it is known that in the R123 system compression of the soft bonds along the *c*-axis leads to an effective charge transfer from the $CuO$ chains to the $CuO_2$ planes. According to Eq. (1) a positive $dn_h/dp$[22] which can be calculated from a simple charge-transfer model (see for example Refs. 19,22,23) results in positive $dT_c/dp$ values in the underdoped and negative values in the overdoped region whereas the pressure effect of optimally doped samples is zero for $\delta p \to 0$, i.e. if pressure only lead to a change of $n_h$, we would expect pressure effects $dT_c/dp \propto (n_h - n_{h,opt})$. Simultaneous compression of the *a*- and *b*-axes, however, causes intrinsic non-charge-transfer effects which almost cancel each other in the optimally doped and overdoped region but lead to large contributions to the hydrostatic pressure effect in the underdoped region[18,19]. From uniaxial pressure and thermal expansion data it is clear that the non-charge-transfer effects in the optimally doped and overdoped region are small and constant. Therefore, for equal $dn_h/dp$ the zero-crossings $dT_c/dp = 0$ of the three systems occur at the same hole concentration close to optimum doping.

The height of the peaks in $T_c/dp$ is possibly influenced by the state of oxygen ordering in the $CuO$ chains. Measurements of $dT_c/dp$ on charge compensated, tetragonal $(Ca_zLa_{1-z})(Ba_{1.75-z}La_{0.25+z})Cu_3O_x$ by Goldschmidt *et al.*[23] did not show any peak like those we found for our orthorhombic samples with chain structure. As it is known that the oxygen order becomes worse with increasing Ca content[24,25] as well as with increasing R-ion radius[17] and, as can be seen in Fig. 2, the height of the peaks decreases with increasing Ca content and increasing R-ion radius, we conclude that the height of the peaks in $dT_c/dp$ is determined by the degree of oxygen ordering. It is much more difficult to explain why the maxima in $dT_c/dp$ occur at a distinct hole concentration $n_{h,m}$. It can be speculated that the particular hole concentration for maximum pressure effect in the different compounds is connected with modifications of energy scales leading to formation and condensation of superconducting pairs due to lattice deformations[26].